\documentclass[letterpaper, 10 pt, conference]{ieeeconf}  

\IEEEoverridecommandlockouts                              
\title{
Task Allocation of UAVs for Monitoring Missions via Hardware-in-the-Loop Simulation and Experimental Validation
}

\author{Hamza Chakraa, François Guérin, Edouard Leclercq and Dimitri Lefebvre
\thanks{The authors are with Université Le Havre Normandie, Normandie Univ, GREAH UR 3220, F-76600 Le Havre, France.
        Email: {\tt\small hamza.chakraa@univ-lehavre.fr}}}

\usepackage{amsmath,amsfonts,amssymb}
\usepackage{algorithm}
\usepackage{algorithmic}
\usepackage{array}
\usepackage[caption=false,font=normalsize,labelfont=sf,textfont=sf]{subfig}
\usepackage{textcomp}
\usepackage{xcolor}
\usepackage{stfloats}
\usepackage{url}
\usepackage{hyperref}
\usepackage{graphicx}
\usepackage{cite}
\let\labelindent\relax
\usepackage{enumitem}
\usepackage{diagbox}   
\usepackage{pifont}    

\begin{document}


\maketitle

\begin{abstract}
This study addresses the optimisation of task allocation for Unmanned Aerial Vehicles (UAVs) within industrial monitoring missions. The proposed methodology integrates a Genetic Algorithms (GA) with a 2-Opt local search technique to obtain a high-quality solution. Our approach was experimentally validated in an industrial zone to demonstrate its efficacy in real-world scenarios. Also, a Hardware-in-the-loop (HIL) simulator for the UAVs team is introduced. Moreover, insights about the correlation between the theoretical cost function and the actual battery consumption and time of flight are deeply analysed. Results show that the considered costs for the optimisation part of the problem closely correlate with real-world data, confirming the practicality of the proposed approach.
\end{abstract}

\begin{keywords}
Multi-robot system (MRS), multi-robot task allocation (MRTA), unmanned aerial vehicle (UAV), industrial monitoring.
\end{keywords}

\section{Introduction}

To avert failures and potential disasters arising from the growth of risky systems and facilities, particularly those involved in manufacturing or storing hazardous chemicals, it is imperative to establish rigorous security measures. Neglecting the protection of vital areas may lead to severe consequences, including the loss of lives and valuable resources. Therefore, an effective strategy for addressing safety concerns seems to be the adoption of advanced technologies, such as artificial intelligence, robotics, and automation. This could involve the integration of advanced sensor technologies alongside a fleet of mobile robots. In practical scenarios, experts usually draft mission plans manually. Hence, using a team of autonomous robots has become a common practice that has been adopted due to its lower cost when compared to manned setups. This process requires giving robots decision-making algorithms so they can divide the work among themselves more quickly, figuring out the data they must collect for the mission and the order in which the tasks should be completed \cite{Chakraa2023a, Chakraa2023b}.

In this research, we explore the challenge of planning inspections within industrial zones by employing a Multi-Robot System (MRS) consisting of Unmanned Aerial Vehicles (UAVs). The issue we're tackling centres on Multi-Robot Task Allocation (MRTA) which aims to efficiently assign tasks to available robots while meeting specific system goals and constraints \cite{Khamis2015, Nunes2017, Chakraa2023}. 

The problem at hand involves managing multiple mobile sensing robots assigned to conduct measurement tasks across various locations within an industrial environment. This paper extends the solution proposed in \cite{Chakraa2023a, Chakraa2023b} by incorporating additional adaptations designed to validate experimentally the proposed task allocation approach using real UAVs. Also, a HIL simulation is presented to provide a controlled and replicable testing environment that closely mimics real-world conditions. By bridging the gap between theoretical modelling and practical implementation, this study aims to present a more resilient and reliable system for distributing UAVs tasks.

The remainder of this paper is structured as follows: the second section provides an overview of the related works about MRTA with real-world applications. Section III presents the problem under study and introduces a mathematical model. Section IV presents the experimental design including UAV control law, communication protocols, and introduces a HIL simulator. Further, Section V presents the results of the simulation and the experimental findings. Finally, Section VI offers a conclusion and suggestions for future research.

\section{Related works}

Patrolling algorithms for multi-robot systems (MRS) have garnered significant attention in both academia and industry \cite{Almadhoun2016}, with the Multi-Robot Task Allocation (MRTA) problem at the core of this research. MRTA can be tackled through Market-based, Behaviour-based, and Optimisation-based methods \cite{Aziz2022}. While many studies have focused on theoretical models and simulations, recent research \cite{Chakraa2023} highlights that few have validated their approaches through real-world experiments. Notable works addressing MRTA with real-world validation include \cite{Parker1998, Werger2000, Gerkey2002, Ponda2010, Wang2023b, Chopra2017, Nekovar2021, Camisa2023}, which are categorized by their approach.

In \cite{Parker1998}, a behaviour-based, fault-tolerant cooperation system for robots in dynamic environments was tested during a hazardous waste cleanup. Similarly, \cite{Werger2000} explored cooperative behaviours using the "Broadcast of Local Eligibility" concept, validated through real-world robot teamwork. \cite{Gerkey2002} introduced auction methods for task allocation, tested in robot bidding scenarios. \cite{Ponda2010} extended a distributed task allocation framework for heterogeneous agents, validated via indoor flight tests. \cite{Wang2023b} proposed the Performance Impact (PI) algorithm for task allocation with deadlines, tested through hardware-in-the-loop experiments.

In optimisation-based approaches, authors \cite{Chopra2017} used a distributed Hungarian method for the Linear Assignment Problem, demonstrated in a multi-robot orchestral performance. Nekovar \textit{et al.} \cite{Nekovar2021} developed inspection methods for UAVs in power line maintenance, tested with real UAVs. And, Camisa \textit{et al.} \cite{Camisa2023} applied Mixed Integer Linear Programming (MILP) to large-scale MRTA for pick-up and delivery, validated through simulation and real-world testing with UGVs and UAVs.

While significant theoretical progress has been made in MRTA, there remains a gap in practical implementation and validation. Bridging this gap through real-world experiments is crucial to ensure the feasibility of proposed methodologies. This work demonstrates the effectiveness of task allocation strategies in industrial environments through HIL simulations and experimental validation.

\section{Optimisation problem formulation}

The mission's environment is presented within a grid-based map of a two-dimensional mesh sized $(N_x \times N_y)$ where $R$ mobile robots $\mathcal{R}=\{r_1,\dots,r_k,\dots,r_R\}$ navigate (a 2D representation is employed, assuming the UAVs maintain a constant flight level). The system is constituted of $V$ cells, identified by the set $\mathcal{V}=\{v_1,\dots,v_j,\dots,v_V\}$ and their central coordinates $(x_j, y_j)$. These cells adhere to specific dimensions based on the problem's needs. Thus, each robot's environment is represented as a weighted graph $\mathcal{G} = (\mathcal{V}, \mathcal{E})$, where $\mathcal{E} : \mathcal{V} \times \mathcal{V} \times \mathcal{R} \rightarrow \mathbb{R}^+$ is the set of edges. Every edge $e \in \mathcal{E}$ connect two vertices $v_j,v_{j'} \in \mathcal{V}$ for a given a robot $r_k$ and holds an elementary cost $\tilde{c}(j,j',k) > 0$. It's essential to note that moving directly from $v_{j}$ to $v_{j'}$ isn't possible if the cells aren't adjacent, in which instance $\tilde{c}(j,j',k) = \infty$. This representation allows for a practical approach to describe the state space, enabling the calculation of costs between positions where robots need to travel while considering obstacles. 

Multiple measurements must be performed within specific locations termed as "sites" within the environment. The set of sites to be visited is represented as $\mathcal{A}=\{a_1,\dots,a_i,\dots,a_A\}$ where $\mathcal{A} \subseteq \mathcal{V}$. The robots start and finish their tour at common site $a_1$ (the depot) where no particular assignment is needed. $\mathcal{M}=\{m_1,\dots,m_q,\dots,m_M\}$ denotes the range of measurement types, likely indicating the diverse data or information the robots need to gather during their mission. The set of tasks that the robots must perform to accomplish the mission is defined as $\mathcal{J} = \{j_1,\dots,j_t,\dots,j_T\}$, where $T$ represents the total number of tasks. In this context, a task $j_t = (i,q)$ is described as a measurement of type $m_q \in \mathcal{M}$ to be executed at a specified site $a_i \in \mathcal{A}$. Additionally, we define the function $\mathcal{T}$: $\mathcal{M} \times \mathcal{A} \rightarrow \{0, 1\}$ in the following way:

\begin{equation*}
    t_{i,q} = \begin{cases}
        1, ~ \text{if} ~ (i,q) \in \mathcal{J} \\ 
        0, ~ \text{otherwise}
    \end{cases}
\end{equation*}

As stated before, a group of $R$ mobile robots that come with distinct sensors for task execution is considered for the mission. These robots vary based on their sensors. Each sensor serves a specific measurement, forming the set $\mathcal{M}$ of all sensors. Each robot $r_k \in \mathcal{R}$ has a set of sensors $\mathcal{M}_k \subseteq \mathcal{M}$. The function $\mathcal{P}$: $\mathcal{M} \times \mathcal{R} \rightarrow {0, 1}$ is defined as:

\begin{equation*}
    p_{k,q} = \begin{cases}
        1, ~ \text{if} ~ m_q \in \mathcal{M}_k \\ 
        0, ~ \text{otherwise}
    \end{cases}
\end{equation*}

Afterwards, the well-known Dijkstra algorithm is applied to compute the minimal cost $c(i,i',k)$ between every two sites $a_i,a_{i'} \in \mathcal{A}$ and for every robot $r_k$ based on the elementary costs defined previously. The objective of this optimisation challenge is to minimise a cost function by effectively allocating robots to tasks and organising sequences of tasks, represented as $\mathcal{S}(k)$ for every robot $r_k$. Subsequently, a cost function must be defined. In the domain of combinatorial optimisation, the MinSum and MinMax global cost functions are widely used. The MinSum objective function $C_1 : \{ \mathcal{S}(k), r_k \in \mathcal{R}\} \rightarrow \mathbb{R}^+$ is utilised to assess the collective cost incurred by all robots. 
Conversely, the MinMax objective function $C_2 : \{\mathcal{S}(k), r_k \in  \mathcal{R}\} \rightarrow \mathbb{R}^+$ optimises the highest individual cost among the robots. The methodology presented in this paper may use both objective functions. Given the context of our study, we consider a typical patrol mission using the MinSum function, and we employ the MinMax function in case of an accident to gather information as quickly as possible. Note that $c(i,i',k)$ can represent either energy (EU) for MinSum optimisation or time units for MinMax optimisation (TU).

In summary, this MRTA problem considers that robots are capable of simultaneously managing either single or multiple tasks in a given site. Each task is managed by one robot within a defined planning period. The problem may be formalised with Mixed Integer Linear Programming (MILP) as detailed in Equations (\ref{eq1}) to (\ref{eq7}). The boolean decision variable $x_{i,j}^k$ is defined such as:

\begin{equation*}
    x_{i,j}^k = \begin{cases}
        1, ~ \text{if} ~ r_k \in \mathcal{R} ~ \text{travel from} ~ a_i \in \mathcal{A} ~ \text{to} ~ a_j \in \mathcal{A} \\ 
        0, ~ \text{otherwise}
    \end{cases}
\end{equation*}

Furthermore, the integer decision variable $u_i^k$ determines the sequence in which robot $r_k$ visits site $a_i$ within its assigned mission. 

Equations (\ref{eq1}) and (\ref{eq2}) define the cost functions: one is accounted for minimising the overall cost of all robots within the MinSum optimisation, while the other aims to minimise the highest cost among the robots in the MinMax optimisation. Equation (\ref{eq3}) specifies that each robot initiates the mission from the depot $a_1$. Equation (\ref{eq4}) mandates that entry into a site must correspond to the exit, ensuring a return to the depot. Equation (\ref{eq5}) eliminates any sub-tours within the sequences of robots, thereby computing directed cycles within each robot’s sequence. Additionally, in addressing the autonomy limitations of the robots, Constraint (\ref{eq6}) imposes that the total mission cost of a robot $r_k$ must not exceed $\mathcal{B}_k$. Lastly, Constraint (\ref{eq7}) guarantees that every task is completed by a robot equipped with the appropriate sensor.

\begin{align}
\text{min} ~~~ & C_1 = \displaystyle  \sum\limits_{r_k \in \mathcal{R}} \sum\limits_{a_i,a_j \in \mathcal{A}} x_{i,j}^k \times c(i,j,k)
\label{eq1} \\
\text{or }& 
\text{min} ~~~ C_2 = \displaystyle \max\limits_{r_k \in \mathcal{R}} \sum\limits_{a_i,a_j \in \mathcal{A}} x_{i,j}^k \times c(i,j,k)
\label{eq2} \\
\text{s.t.} ~~~ & \displaystyle \sum\limits_{a_i\in \mathcal{A}} x_{1,i}^k = 1 ~~~~~~~~~~~~~~~~~~~~~~~~~~~~~ \forall r_k \in \mathcal{R} 
\label{eq3} \\
            & \displaystyle \sum\limits_{a_j \in \mathcal{A}}  x_{i,j}^k = \sum\limits_{a_j \in \mathcal{A}} x_{j,i}^k ~~~~~~~ \forall a_i \in \mathcal{A},~~ \forall r_k \in \mathcal{R} 
\label{eq4} \\
            \begin{split}
            & \displaystyle u_{i}^k + x_{i,j}^k \leq u_j^k + (A -1) \times (1-x_{i,j}^k) \\
            &~~~~~~~~~~~~~~ \forall a_i \in \mathcal{A},~ \forall a_j \in \mathcal{A} \slash \{a_1\},~ \forall r_k \in \mathcal{R}
            \end{split}
\label{eq5} \\
            & \displaystyle\sum\limits_{a_i,a_j \in \mathcal{A}} x_{i,j}^k \times c(i,j,k) \leq \mathcal{B}_k ~~~~~~~~~~~\, \forall r_k \in \mathcal{R}
\label{eq6}\\
            \begin{split}
            & \displaystyle\sum\limits_{r_k \in \mathcal{R}}\sum\limits_{a_i \in \mathcal{A}} p_{k,q}  \times x_{i,j}^k \geq t_{j,q} \\
            &~~~~~~~~~~~~~~~~~~~~~~~~~~~~~~~ \forall m_q \in \mathcal{M},~ \forall a_j \in \mathcal{A}
            \end{split}
\label{eq7} 
\end{align}

\section{Optimisation methodologies}

In order to solve this problem, Genetic Algorithms (GA) are proposed. The solution is modelled as a three-line table. The first line is for sites where tasks are to be executed by robots. The second line displays measurements linked to these sites, while the third line designates the assigned robots. Each column in this table, comprising a site, a measurement, and an assigned robot, represents a gene. The GA begins with initialising a population, creating the initial set of individuals across $\tau_g$ generations, where each generation contains $\tau_p$ individuals. The initial population is randomly generated, considering only suitable robots for tasks to avoid unfit solutions. Afterwards, selection and recombination will drive the population towards better solutions. In this study, crossover swaps robot assignments between corresponding genes and modifies task routing within chromosomes. Meanwhile, mutation randomly changes robot assignments for tasks. For further details about the approach, readers are directed to \cite{Chakraa2023a}.


Given the complexity of the problem at hand, solutions derived from GA may yield less-than-optimal outcomes. To improve the quality of these solutions, we implement the 2-Opt local search technique. Its primary objective is to refine the routing component within the solutions obtained from GA. By employing 2-Opt as a post-processing step for each robot's sequence, we aim to augment the efficiency and optimality of the routing scheme, ultimately enhancing the overall performance of our task allocation system. For further details about the approach, readers are directed to \cite{Chakraa2023b}.

\section{Experimental setup}

\begin{figure*}[t]
   \centering
   \def\svgwidth{0.9\textwidth}
    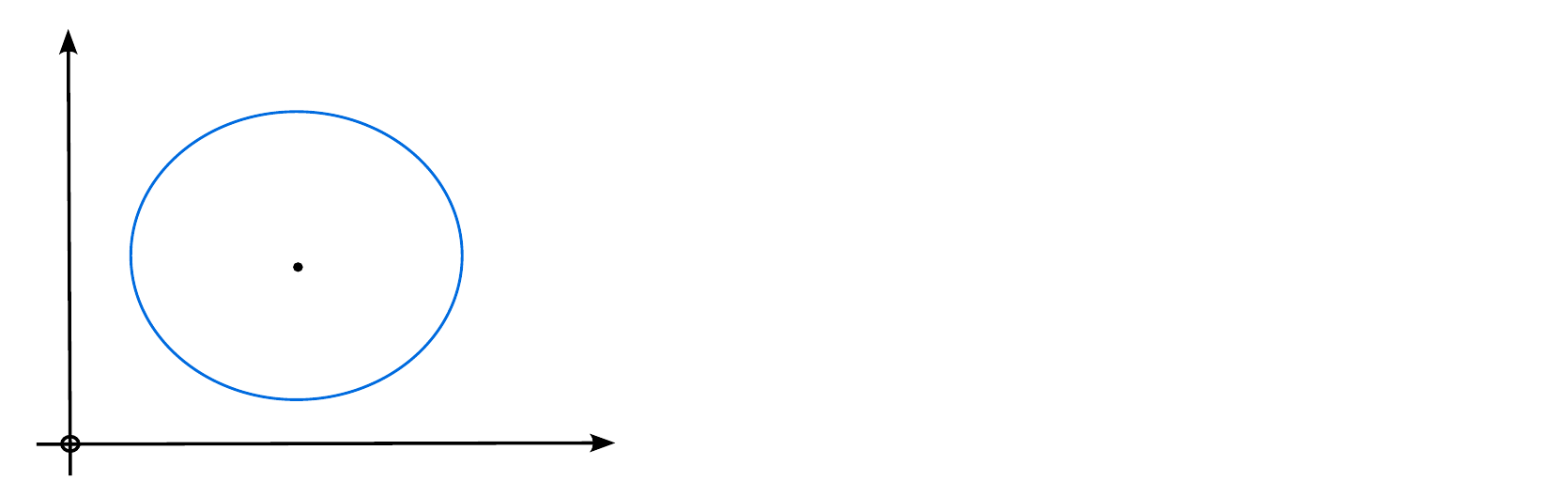
   \caption{UAV system and target description. (a) Utilising Fourier Descriptors for site characterisation (circular shape) with the centre $C$ as a reference point. (b) Defining the architecture and the position errors of the UAV.}
    \label{1}
\end{figure*}

This section presents the experimental setup of this study including the control law governing the UAV fleet operations, the collision avoidance mechanism, and the communication protocols among drones. Furthermore, the section will introduce a HIL simulator that reproduce the behaviour of the drones.
Finally, a specific task allocation scenario is introduced to illustrate the solution generated by GA and 2-Opt for the UAV fleet.

\subsection{Control framework of the UAVs}

Figure \ref{1} presents the UAV architecture in order to develop the control laws. Here, our aim is for the UAV to locate the target point $M$ using only the shape's reference point $C$. To achieve this, our functions are designed to conform to a cyclic period of $2\pi$ in polar coordinates from the reference point. These functions can be sampled, and rotational quaternions are provided to enable 3D rotation of the planar shape by the UAV. It must continuously minimise the distance errors from the target point $M$ to approach zero. Simultaneously, it is required to orient towards the reference point $C$.

\begin{equation}
    d_{\beta} = \sqrt{X_{\beta}^2+Y_{\beta}^2}
    \label{8}
\end{equation}

The Fourier descriptors describe a planar function $d_\beta$ (Equation (\ref{8})) with a single variable $\beta$ from a reference point $M$ (Figure \ref{1}(a)). In this study, we use the centroid distance to define the signature shape $d_\beta$ of the desired curve, as it effectively captures the complex patterns with the highest degree of precision. It is defined as the distance between the boundary point $M$ and the centre $C$ of the shape. To align with our research context, reference point $C$ corresponds to the centre of the site $a_i$ under consideration. The distance $d_{\beta}$ is computed with the bearing angle $\beta$ (the angle between the $X$-axis and the direction defined by point $M$ and the reference point $C$) \cite{Petitprez2023}. \\

\subsubsection{Control Law}

The control inputs for the UAVs are roll ($\Phi$), pitch ($\Theta$), yaw ($\Psi$), and the vertical acceleration ($A_Z$). The principal aim of the cascading controller strategy proposed here is to regulate the UAV's position in relation to a stationary target point while ensuring the maintenance of safety standards. This is achieved through the implementation of two distinct controllers. The velocity controller evaluates the velocity errors along the $x$, $y$, and $z$ axes. A saturation function is introduced to constrain the pitch/yaw angles and vertical acceleration, activating when the cumulative speed error surpasses a predetermined threshold. 
The outputs of the velocity controller serve as the reference inputs for the UAV's flight controller. The position controller (outer loop) considers the positional errors along the $x$, $y$, and $z$ axes. A secondary saturation function is employed to regulate the velocity concerning the $x$, $y$, and $z$ axes when the position error exceeds a specified limit. Under this limit, the UAV adjusts its position towards the target; otherwise, it proceeds at a fixed maximum speed. The outputs of the position controller are directed as inputs to the velocity controller. \\

\subsubsection{Velocity Regulation}

The filtered velocity errors $\hat{\varepsilon}_{v}$ for the UAV are as follows:

\begin{align}
\hat{\varepsilon}_{v} : \left\{ 
\begin{array}{l}
\dot{e}_{vx} = V_{x} - \hat{V}_{x} \\
\dot{e}_{vy} = V_{y} - \hat{V}_{y} \\
\dot{e}_{vz} = V_{z} - \hat{V}_{z}
\end{array}
\right.
\end{align}

The goal for velocity control is given by:

\begin{align}
\lim_{t \to \infty} \hat{\varepsilon}_{v}(t) = 0
\end{align}

To achieve the control objective, the following velocity control law has been applied:

\begin{equation}
\begin{bmatrix}
\dot{\hat{e}}_{vx} \\
\dot{\hat{e}}_{vy} \\
\dot{\hat{e}}_{vz}
\end{bmatrix} =
\begin{bmatrix}
f \left(\sum_{j=1}^{3} kv_{xj} \epsilon v_{xj}, \bar{\Theta}, \bar{V}_{x}, 0\right) \\
f \left(\sum_{j=1}^{3} kv_{yj} \epsilon v_{yj}, \bar{\Phi}, \bar{V}_{y}, 0\right) \\
f \left(\sum_{j=1}^{3} kv_{zj} \epsilon v_{zj}, \bar{A}_{z}, \bar{V}_{z}, g\right)
\end{bmatrix} = \vec{V}_{\text{CMD}}
\end{equation}

Note that the coefficients $kv_{xj}$, $kv_{yj}$, $kv_{zj}$ are to be adjusted, $\bar{\Theta}$, $\bar{\Phi}$, $\bar{A}_{z}$ are the maximal angular positions and vertical acceleration that can be reached when the velocity errors exceed $\bar{V}_{x}$, $\bar{V}_{y}$, $\bar{V}_{z}$. The command vector of the UAV is $\vec{V}_{\text{CMD}} = (\bar{\Theta}, \bar{\Phi}, \bar{\Psi}, \bar{A}_{z})^T$. $f$ is a saturation function. \\

\begin{figure*}[h]
    \centering
    \def\svgwidth{1\textwidth}
    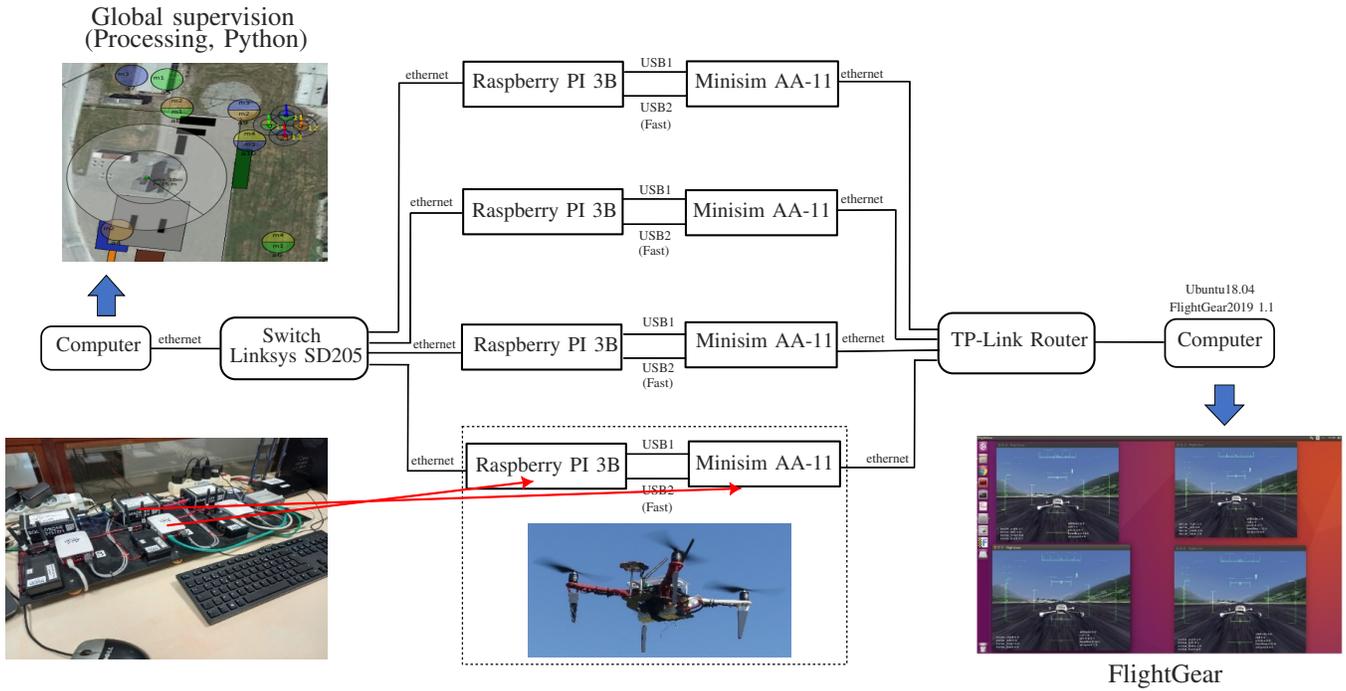
    \caption{\centering Detailed overview of the simulator device and its mission supervision interfaces}
    \label{2}
\end{figure*}

\subsubsection{Position Adjustment}
To align with the designated orientation $\beta$ determined through Fourier descriptors, the control for yaw is adjusted to match $\beta$. The position errors $\hat{\epsilon}_{p1}$ for the UAV are defined as:

\begin{equation}
\hat{\epsilon}_{p}:
\left\{
\begin{aligned}
&\hat{e}_{px} = \hat{d} \cos (\beta - \hat{\psi}) \cos(\gamma) - d_{\beta} \\
&\hat{e}_{py} = \hat{d} \sin (\beta - \hat{\psi}) \sin(\gamma) \\
&\hat{e}_{pz} = (Z_c - d_{\beta} \sin (\gamma)) - Z
\end{aligned}
\right.
\end{equation}

with

\begin{equation}
\lim_{t \to \infty} \hat{\epsilon}_{p}(t) = 0
\end{equation}

To satisfy this position control goal, the position control strategy applied is:

\begin{equation}
\begin{bmatrix}
\dot{\hat{e}}_{px} \\
\dot{\hat{e}}_{py} \\
\dot{\hat{e}}_{pz}
\end{bmatrix} =
\begin{bmatrix}
f \left(\sum_{j=1}^{3} kp_{xj} \epsilon p_{xj}, \bar{V}_{x}, \bar{D}_{x}, 0\right) \\
f \left(\sum_{j=1}^{3} kp_{yj} \epsilon p_{yj}, \bar{V}_{y}, \bar{D}_{y}, 0\right) \\
f \left(\sum_{j=1}^{3} kp_{zj} \epsilon p_{zj}, \bar{V}_{z}, \bar{D}_{z}, 0\right)
\end{bmatrix} = \vec{V}
\end{equation}

Note that the parameters $kp_{xj}, kp_{yj}, kp_{zj}$ are to be tailored. $\Vec{V} = (V_x, V_y, V_z)^T$ is the linear velocity vector of the UAV with $V_x, V_y$ and $V_z$ the maximal linear velocities reached when the position errors are above the distances $\bar{D}_{x}$, $\bar{D}_{y}$, and $\bar{D}_{z}$. For more details, readers are directed to \cite{Petitprez2023}. \\

\subsubsection{Collision Avoidance mechanism}

The developed avoidance mechanism allows for avoiding both stationary and moving UAVs by providing real-time GPS coordinates. It dynamically calculates a repulsive velocity that influences the UAV's trajectory based on the proximity to obstacles or other drones, ensuring safe navigation. This proposed method conforms to the Artificial Potential Field (APF) technique, where the repulsion generated between robots simulates a repulsive potential field. 

\begin{align}
D_{ij} &= \sqrt{d_{ij}^2 + (Z_j - Z_i)^2}
\label{15}
\end{align}

When examining two robots $r_i$ and $r_j$, we represent the distance between them as $D_{ij}$. It's worth noting that $D_{ij}$ can be calculated from GPS coordinates, employing Haversine formulas \cite{Robusto1957}. By setting a maximum distance threshold, denoted as $\Bar{D}_j$, robots can define a safeguarded spherical zone around themselves. When another robot reaches this predetermined safety perimeter, causing $D_{ij}$ to dip below the threshold, the repulsive response is triggered. 
The distance $D_{ij}$ is defined in Equation (\ref{15}) where $d_{ij}$ is the horizontal distance between $r_i$ and $r_j$, and $Z_i$, $Z_j$ are their respective altitudes.




\begin{equation}
\|\vec{V}_{ij}\| =
\left\{
\begin{array}{ll}
\bar{V}_j \cos \left(\frac{\pi D_{ij}}{2\Bar{D}_j}\right) & \text{if } D_{ij} < \Bar{D}_j, \\
0 & \text{otherwise}.
\end{array}
\right.
\label{eq16}
\end{equation}

The intensity of the repulsive effect $\|\vec{V}_{ij}\|$ (Equation \ref{eq16}) is determined based on the distance between the UAVs, and activated when $D_{ij}$ is less than the predefined threshold $\Bar{D}_j$ where $\bar{V}_j$ represents the maximum intensity of repulsive velocity for robot $r_j$, which is an adjustable parameter \cite{Petitprez2023}.


\subsection{Communication aspects and simulation system}

The drones are connected within a "mesh" network with 2.4 GHz "Microhard pMDDL2450-LC" radio modules. In this wireless network topology, all drones are connected peer-to-peer without a central hierarchy and are capable of receiving, sending, or relaying messages with the UDP multi-cast protocol. If a drone leaves or joins the network, the others automatically adapt to the new configuration. This allows for sharing especially GPS coordinates so the distances between the robots are calculated in a real-time manner and used for the collision avoidance framework.

Afterwards, in order to simulate a specific mission, a software application has been developed in Processing. It allows for the real-time visualisation of drone movements, as their GPS coordinates are transmitted via UDP multi-cast protocol frames (Figure \ref{2}). This simulation system provides a highly accurate and realistic representation of the physical drones, enabling precise simulation of their behaviours and interactions. By using HIL simulation, potential issues can be identified and addressed in the virtual model before deploying physical drones, reducing the risk of costly errors and also allow to supervise the mission of the drones during the experiments. This simulation system consists of:

\begin{itemize}
    \item Four hardware-in-the-loop \textit{MiniSim} simulators (\href{https://squadrone-system.com}{Squadrone Systems}). Each of them reproduces the dynamic behaviour of the UAV and its flight controller (Figure \ref{2}).
    \item Four \textit{Raspberry PI 3B+} companion computers, connected to the \textit{MiniSim} simulators and on which our algorithms were implemented in C++ with an open API (\href{https://squadrone-system.com}{Squadrone Systems}). These computers were exactly the same installed on real UAVs.
    \item A laptop, connected (by a switch or by radio modules 2.4GHz to emulate the real “mesh” connection) to the \textit{Raspberry PI 3B+} companion computers, for programming and starting the execution of the algorithms.
    \item An open-source flight simulator \href{https://www.flightgear.org}{Flightgear} (v. 2020.3) installed on a separate computer and connected to the \textit{MiniSim} simulators via a rooter to visualise the flight of the UAVs. This setup allowed several sessions to be run simultaneously in order to observe the flight of the group (one UAV per window, as illustrated on Figure \ref{2}).
    \item 
    A supervision tool has also been designed (Processing, Python) to get a global view of the team of UAVs with sites marked with circles, measurements represented with distinct colours, and obstacles shown with transparent circles (Figure \ref{2}).
\end{itemize}

\subsection{The considered scenario}

To demonstrate the effectiveness of the proposed approach, an area in the Laugui Concept Fire Engineering Consultancy site (Normandy, France) was selected for our experiments. The developed simulator (Figure \ref{2}), uses the Laugui Concept Area for a global visualisation and supervision of the considered scenario. Initially, the area was converted into a 2D grid map to calculate travel costs between various sites and determine the optimal solution. Measurement tasks were assigned to ten distinct locations, denoted as $\mathcal{A} = \{a_1, \dots, a_{10}\}$, with $a_1$ representing the depot. These tasks included various activities and were divided into four distinct types of measurements, denoted as $\mathcal{M} = \{m_1, m_2, m_3, m_4\}$. Each location $a_i$ was assigned a subset of measurement types from $\mathcal{M}$, except for $a_1$ where no measurement was required.

The team consists of four robots, $\mathcal{R} = \{r_1, r_2, r_3, r_4\}$, each equipped with different sensors capable of performing the measurements $\mathcal{M}$. Specifically, robot $r_1$ is equipped with sensors for $m_1$ and $m_3$, robot $r_2$ have a sensor for $m_2$, robot $r_3$ have sensors for $m_2$ and $m_3$, and robot $r_4$ could perform $m_1$ and $m_4$. 
Additionally, the scenario is composed of $15$ tasks, $\mathcal{J} = \{j_1, \dots, j_{15}\}$. 



\section{Experiments}

In this section, we present the results of the conducted experiments to evaluate the performance of our optimisation methodologies. Note that, only real-world results from UAVs are presented because since the focus is on validating the methodologies in real-world scenarios to ensure their practical applicability, so the simulator is used primarily as a supervision tool in this study. We consider two experiments: one with the MinSum objective function and another with the MinMax objective. The mission is determined using the combination of GA and 2-Opt local search, which optimises the sequence of site visits and the overall cost of the mission. 

\begin{figure}[h]
    \centering
    \def\svgwidth{0.32\textwidth}
    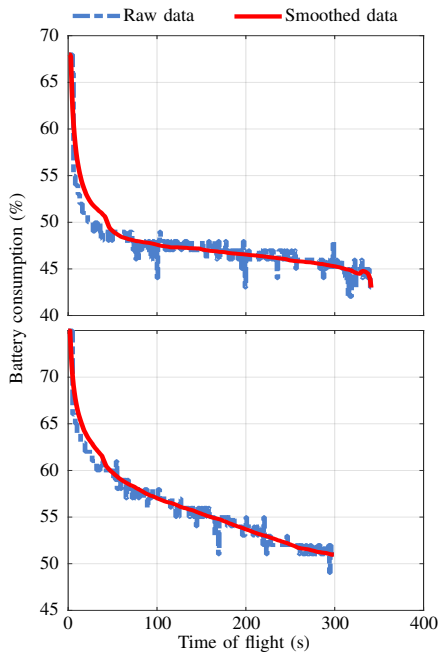
    \caption{Energy consumption for the two robots $r_3$ and $r_4$ during the mission}
    \label{3}
\end{figure}

For the MinSum optimisation, although four robots are available, the optimal solution requires only robots $r_3$ and $r_4$ to complete all the tasks and optimise the objective function. Subsequently, the first experiment was conducted using two real drones. It aimed to analyse the correlation between battery usage and the cost function, particularly the assumption that battery usage is correlated with the cost of the MinSum function and correlated with time for MinMax optimisation.

\begin{figure*}[h]
    \centering
    \def\svgwidth{0.7\textwidth}
    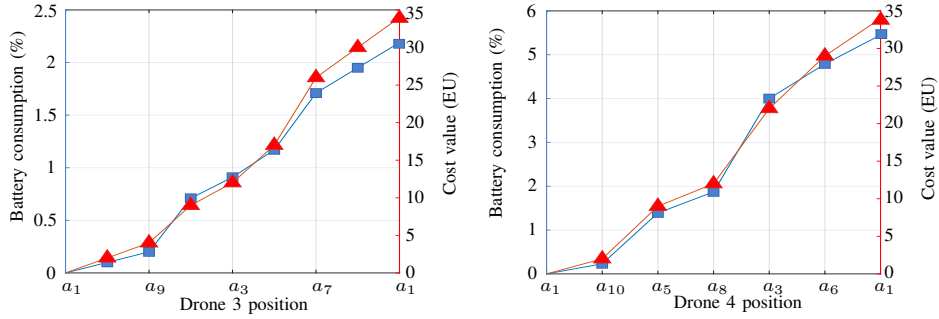
    \caption{Energy consumption and cost between each two sites in the sequence of robots $r_3$ and $r_4$ for the MinSum optimisation}
    \label{4}
\end{figure*}

\begin{figure*}[h]
    \centering
    \def\svgwidth{1.06\textwidth}
    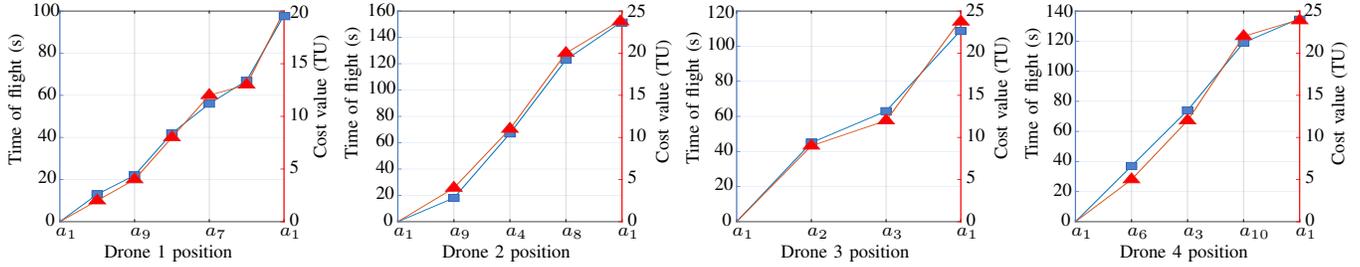
   \caption{Time of flight and cost between each two sites in the sequence of drones $r_1$, $r_2$, $r_3$, and $r_4$ for the MinMax optimisation}
    \label{5}
\end{figure*}

Figure \ref{3} shows the energy consumption for both UAVs used for the solution obtained with MinSum optimisation. The energy consumption data is plotted over time, indicating the energy consumption as the robots execute their assigned tasks. The data points, shown in blue, represent the instantaneous battery consumption, while the red lines indicate the smoothed data, providing a clearer view of the overall consumption trends and filtering out the noise from the raw data. Both UAVs show an initial rapid drop in battery percentage within the first 50 seconds, due to the high power demand during take-off and initial stabilisation. Following this initial phase, the rate of battery consumption stabilises, showing a more gradual decline. Both UAVs exhibit approximately similar consumption profiles, with minor differences in their exact rates.

Figure \ref{4} illustrates the battery consumption and cost incurred by robot $r_3$ and $r_4$. It presents the cumulative battery consumption and cumulative cost at each step of the mission, offering an overview of the total battery usage and overall cost as the robot progresses through the visited sites. The data between battery consumption and the costs for robots $r_3$ and $r_4$, as illustrated in Figures \ref{4} (left) and \ref{4} (right), reveals a strong correlation. This correlation indicates that as the battery consumption increases, the associated cost also rises in an approximately similar manner. As each robot progresses through its mission, the total battery consumption and the overall cost grow concurrently. This parallel development implies that the cost function effectively captures the energy expenditure of the robots. The close alignment of the curves in these graphs validates the cost model used in the mission planning.

Similarly, the same experiments were considered for the generated solution with the MinMax objective function. Figure \ref{5} illustrates the time of flight and cost incurred by the four robots throughout the mission. This figure shows a strong correlation between the time and the cost calculated as in each movement the time spent is proportional to the travelling cost. In summary, the considered configuration of mesh with an elementary cost that equals one between every two cells is adapted to the real-world scenario and considers consumed time and energy.

The slight differences in results may be attributed the simplification in the mesh configuration as it might not perfectly capture real-world complexities, leading to minor disparities. Variability in real-world factors such as weather conditions (e.g, wind) can also contribute to these differences, resulting in the observed slight variations.

\section{Conclusion}

This paper presents a MRTA problem with an application using a fleet of UAVs. Our developed HIL simulator accurately models the real-world behaviour of UAVs in monitoring missions. The analysis of battery usage during the experiment, and the modelled cost function confirms the effectiveness of our task allocation and optimisation methodologies. Looking ahead, online and decentralised approaches present promising avenues for enhancing the capabilities of robots in monitoring missions. By implementing real-time decision-making strategies, UAVs can adapt to dynamic environments more effectively, adjusting their tasks and routes based on live data inputs. Furthermore, our system can be used as a digital twin, facilitating the combination of real and virtual drones. This integration allows for the exchange of information between physical UAVs and their virtual counterparts. This can lead to better calibration of the simulation models and more accurate future predictions. Also, integrating machine learning techniques can enhance the decision-making capabilities of UAVs. For instance, reinforcement learning algorithms can enable UAVs to learn from their experiences and improve their task allocation strategies over time.

\bibliographystyle{unsrt}
\bibliography{references}

\end{document}